\documentclass[11pt]{article}

\usepackage{natbib}
\usepackage{hyperref}

\usepackage{aas_macros}

\newcommand{\vmax}{$v_\mathrm{max}$ }
\newcommand{\vpeak}{$v_\mathrm{peak}$ }

\newcommand{\Ms}{ \mathrm{M}_\odot }

\newcommand{\LCDM}{$\Lambda$CDM }

\def\aap{A\&A}                
\def\mnras{MNRAS}            
\def\apj{ApJ}                 
\def\apjl{ApJL}                
\def\nat{Nature}              
\def\pasa{PASA}             
\def\araa{ARA\&A}            

\usepackage{xcolor}
\usepackage{graphicx}

\title{The Milky Way's plane of satellites: consistent with \LCDM}

\author{Till Sawala$^{1\ast}$,
Marius Cautun$^{2}$,
Carlos S. Frenk$^{3}$,
John Helly$^{3}$, \\
Jens Jasche$^{4}$, 
Adrian Jenkins$^{3}$,
Peter H. Johansson$^{1}$,
Guilhem Lavaux$^{5}$, \\
Stuart McAlpine$^{1, 4}$,
Matthieu Schaller$^{2}$ \\
\\
\scriptsize{$^{1}$Department of Physics, University of Helsinki, 00560 Helsinki, Finland,} \\
\scriptsize{$^{2}$Lorentz Institute for Theoretical Physics, Leiden University, Leiden NL-2300, The Netherlands,}\\
\scriptsize{$^{3}$Institute for Computational Cosmology, Durham University, Durham DH13LE, United Kingdom,}\\
\scriptsize{$^{4}$The Oskar Klein Centre, Department of Physics, Stockholm University, Stockholm 106 91, Sweden,}\\
\scriptsize{$^{5}$CNRS \& Institut d'Astrophysique de Paris, Sorbonne Universit\'e, 75014 Paris, France}\\
\scriptsize{$^\ast$To whom correspondence should be addressed; E-mail:  till.sawala@helsinki.fi.}
}

\date{}

\begin{document} 

\maketitle

\begin{abstract}
{\bf The “plane of satellites problem” describes the arrangement of the Milky Way’s 11 brightest satellite galaxies in a remarkably thin plane, possibly supported by rotation. This is in apparent contradiction to the standard cosmological model, wherein the Galaxy is surrounded by a dispersion-supported dark matter halo. Here, we show that the reported exceptional anisotropy of the satellite system is strongly contingent on a lopsided radial distribution, which earlier simulations have failed to reproduce, combined with the close but fleeting conjunction of the two most distant satellites, Leo I and Leo II. Using Gaia proper motions, we show that the orbital pole alignment is much more common than previously reported, and reveal the plane of satellites to be transient rather than rotationally supported. Comparing to new simulations, where such short-lived planes are common, we find the Milky Way satellites to be compatible with standard model expectations.}
\end{abstract}

\section{Introduction}
A key prediction of the standard $\Lambda$ Cold Dark Matter (\LCDM) cosmological model \citep{Davis-1985} is that galaxies such as the Milky Way (MW) are surrounded by a dark matter halo and by satellite galaxies formed within its substructures. In apparent contradiction, the ``plane of satellites" describes the arrangement of satellite galaxies in a thin \citep[e.g.][]{Lynden-Bell-1976, Kroupa-2005, Pawlowski-2018}, possibly rotationally supported \citep{Metz-2008} plane.

Whereas energy dissipation can lead to the gas inside of galaxies settling into rotating thin disks, such a configuration is highly unlikely to form from the collisionless dark matter halo. While other apparent discrepancies between predictions and observations of Milky Way satellites have been resolved through baryonic effects \citep{Navarro-1996, Pontzen-2014, Sawala-2016a} there is no plausible formation mechanism for rotating satellite planes within dispersion-supported dark matter halos. Consequently, the ``plane of satellites problem" has emerged as the most persistent challenge to the dark matter paradigm \citep{Kroupa-2012, Bullock-2017, Perivolaropoulos-2021}.

That the ``plane of satellites" problem has so far eluded resolution
is not for lack of trying. Planes of satellites were found with the same (low) frequency in collisionless and hydrodynamic cosmological simulations \citep{Cautun-2015, Ahmed-2017, Muller-2021} and in MW analogues in isolation or in pairs \citep{Forero-Romero-2018, Pawlowski-2019b}; planes show no significant correlation with other properties of the host halo \citep{Pawlowski-2019b}. There is evidence that filamentary accretion \citep{Libeskind-2005, Shao-2018} or the presence of massive satellites \citep{Samuel-2021} can generate some anisotropy, but systems as thin as the Milky Way's are still very rare \citep{Cautun-2015}. Moreover, any planes that do form in \LCDM are transient, chance alignments of substructures \citep{Muller-2021, Buck-2016, Shao-2019, Samuel-2021}, rather than long-lived, rotationally supported disks.

With no apparent explanation within \LCDM, the ``plane of satellites" might constitute evidence for MOND \citep{Famaey-2012}, an entirely different cosmological framework, in which the Milky Way's satellite galaxies are dark-matter-free ``tidal" dwarf galaxies formed during a hypothetical past close encounter between the MW and M31 \citep{Yang-2014, Bilek-2018, Banik-2018}.

Here, we re-examine the contention that the MW contains an exceptional plane of satellites, explain the origin of the observed anisotropy, and study its time evolution in light of proper motion measurements by the {\it Gaia} space telescope\citep{McConnachie-2020}.

This paper is organised as follows. We define the metrics for spatial and orbital anisotropy in Section~\ref{sec:methods:spatial} and~\ref{sec:methods:look-elsewhere}, and introduce the Gini coefficient mechanism that captures the dependence of the anisotropy on the radial distribution in Section~\ref{sec:methods:gini}. We describe the observational data and its analysis in Sections~\ref{sec:methods:observations}--\ref{sec:methods:integration}, and our \LCDM simulations in Sections~\ref{sec:methods:simulations} -- ~\ref{sec:methods:orphans}. Our results for the Milky Way's plane of satellites are shown in Section~\ref{sec:results}. We analyse the spatial anisotropy in Section~\ref{sec:spatial}, the orbital anisotropy in Section~\ref{sec:orbital}, and the time evolution in Section~\ref{sec:evolution}. We conclude with a summary in Section~\ref{sec:summary}.

\section{Methods}

\subsection{Definition of spatial anisotropy}\label{sec:methods:spatial}
The Milky Way's ``plane of satellites" canonically consists of the 11 ``classical" satellites, the brightest within $r=300$~kpc of the Galactic centre, believed to constitute a complete sample. To characterise the spatial anisotropy of a satellite system, it is customary to consider the inertia tensor, defined as
\begin{equation}
I_{ij} = \sum_{n=1}^N x_{n,i} x_{n,j},
\label{eqn:inertia}
\end{equation}
where $x_n$ are the coordinates of the $n$-th satellite relative to the centre of positions. We label the square roots of its eigenvalues as $a$, $b$ and $c$, corresponding to the dispersions in position along the unit eigenvectors, $\vec{x_a}$, $\vec{x_b}$ and $\vec{x_c}$. A related metric is the ``reduced" inertia tensor defined after projection of the positions onto a unit sphere. We label the square roots of its eigenvalues as $a_\mathrm{red}$, $b_\mathrm{red}$ and $c_\mathrm{red}$. Both $c/a$ and $(c/a)_\mathrm{red} \equiv c_\mathrm{red}/a_\mathrm{red}$ parametrise the spatial anisotropy, smaller values imply greater anisotropy. Note that for small $N$, the expectation values of $c/a$ and $(c/a)_\mathrm{red}$ decrease, regardless of the underlying anisotropy \cite{Santos-Santos-2020}.

\subsection{Definition of orbital anisotropy} \label{sec:methods:look-elsewhere}
To characterise the clustering of orbital poles, we adopt the orbital pole dispersion for a subset of $N_{s}$ satellites, $\Delta_{std}$, defined by \cite{Pawlowski-2019a} as:
\begin{equation}
    \Delta_{\rm{std}} (N_{s}) = \sqrt{ \frac{1}{N_{s}} \sum_{i=1}^{N_{s}} \theta_i^2 },
\end{equation}
where $\theta_i$ is the angle between the orbital pole of the $i$th satellite and the mean orbital pole of the satellites in the subset. To compute the clustering of an observed system relative to expectation, the same analysis is performed on the observations and simulations.

Based on earlier {\it Gaia} DR2 and HST proper motions, \cite{Pawlowski-2019a} calculated orbital pole dispersions for all possible satellite subsets in the MW and in simulations with $N_{\rm{subset}}=3 ... 11$, and discovered that $N_s = 7$ yielded the most unusual configuration. However, there is no a priori reason to specifically consider $N_s=7$. When considering only a proper subset of satellites, the interpretation of $\Delta_{\rm{std}} (N_{s})$ as evidence for unusual clustering is subject to the ``look elsewhere effect''. To account for this, we follow here the method of \cite{Cautun-2015}, which involves performing the same analysis for the simulated systems. As in the observations, we consider all subsets of size $N_s=3 ... 11$ in each simulated system, and identify the most unlikely to arise by chance from an isotropic distribution, which we calculate based on $10^5$ isotropic distributions of $N=11$ points, and the probability distributions of $\Delta_{\rm{std}} (N_s)$ for all $N_s=3 ... 11$ possible subsets.

\subsection{The Gini coefficient of inertia} \label{sec:methods:gini}
As each satellite contributes to the inertia (Equation~\ref{eqn:inertia}) proportional to $r_i^2$, $c/a$ is sensitive to the radial profile. To quantify this relationship, we introduce the Gini coefficient formalism. The central panel of Figure~\ref{fig:radial} shows the summed weights of the closest $i$ satellites from the centre, $\sum\limits_{j=1}^{i}r_j^2$, normalised by the total weight of all 11 satellites, $\sum\limits_{j=1}^{11}r_j^2$. The area between each curve and the diagonal measures the inequality of the satellites' contributions to the inertia, or the sample {\it Gini coefficient of inertia},

\begin{equation}
 G = \left. \frac{1}{N-1} \sum\limits_{i=1}^N (2i - N -1) r^2_i \middle/ \right.  \sum\limits_{i=1}^{N}r^2_i . \label{eqn:gini}
\end{equation}

\begin{figure}[ht]
    \includegraphics[width=\columnwidth]{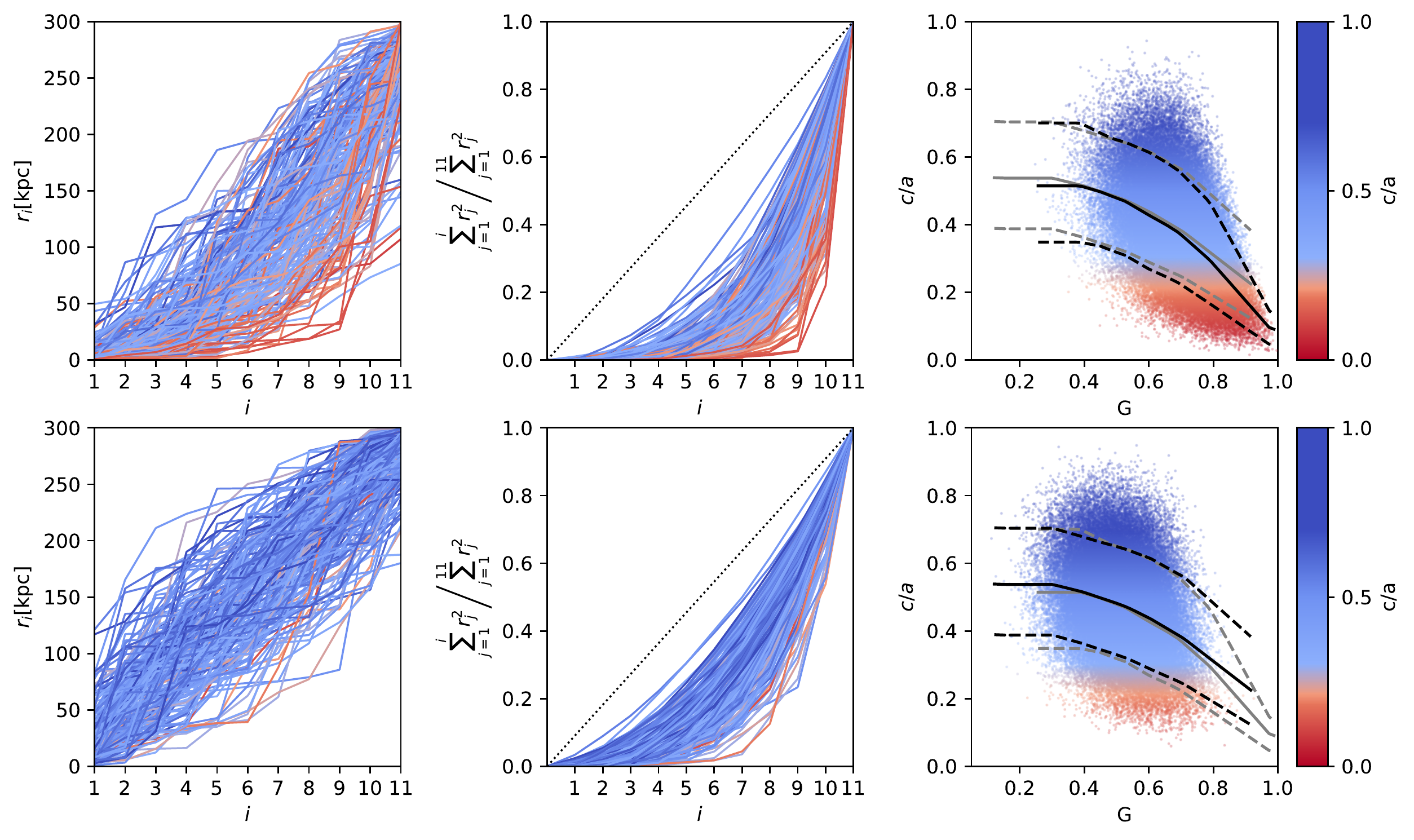}
    \caption{Relation between Gini coefficient of inertia, $G$, and
    anisotropy, $c/a$, for $10^5$ random samples of 11 points, drawn
    from isotropic angular distributions and radial distributions
    uniformly distributed in $r^{1/2}$ (top) or $r$ (bottom). On the right panel, black lines denote the $90^\mathrm{th}$, $50^\mathrm{th}$ and $10^\mathrm{th}$ percentiles of each dataset, grey lines repeat the corresponding percentiles for the other dataset. For clarity, the left
    and middle panels only show the first 200 samples. }
    \label{fig:radial-synthetic}
\end{figure}

Figure~\ref{fig:radial-synthetic} illustrates the relation between the radial distribution and the Gini coefficient, $G$, and the correlation between the $G$ and the measured anisotropy $c/a$, for samples of 11 points drawn from isotropic angular distributions. A more centrally concentrated radial distribution (higher $G$, as shown in the top panel) to greater anisotropy (lower $c/a$).

Compared to a more equal distribution, the Milky Way's centrally concentrated radial profile is equivalent to sampling a system with fewer points. For the purpose of computing $c/a$, the effective sample size \cite{Kish-1965} is only $N_\mathrm{eff}=4.16$.

\subsection{Observations} \label{sec:methods:observations}
We adopt the sky positions and radial velocities from the \cite{McConnachie-2012} catalogue, and combine these, where available, with the \cite{McConnachie-2020b} proper motion measurements based on {\it Gaia} EDR3 \citep{McConnachie-2020a}. The systemic proper motions were measured within a Bayesian framework that combines information from stars with full astrometric data with information from stars with only photometric and proper motion data. The method is a mixture model that associates a probability for each candidate star to be associated with the target galaxy taking into account foreground and background contaminants. For the three innermost satellites (Sagittarius dSph, the LMC and the SMC), where the \cite{McConnachie-2020b} catalogue does not include proper motions, we use the {\it Gaia} DR2 proper motions of \cite{Riley-2019}. We further compiled the most recent estimates of the distance moduli of each satellite. The distance moduli, sky coordinates, radial velocities and proper motions used in this study, including their sources, are listed in Table~\ref{tab:observations}. As discussed in Appendix~\ref{appendix:gaia}, we also repeated our analysis using the {\it Gaia} EDR3 proper motions of \cite{Battaglia-2022}, and using only the {\it Gaia} DR2 proper motions described in \cite{Riley-2019}.

We account for measurement errors by generating Monte Carlo samples of the satellites in the space of observed quantities: distance modulus, radial velocity and proper motions, as well as the position of the Sun relative to the Galactic centre. We model each observable as a Gaussian distribution with the mean value and standard deviation given by the measurements and their quoted errors. For the sun's distance from the Galactic centre, we assume $R_\odot= (8.178 \pm 0.022) ~\rm{kpc}$ \citep{Gravity-2019}, for the circular velocity at the Sun's position, $V_{\rm circ} = (234.7\pm1.7)~\rm{km/s}$ \citep{Nitschai-2021}, and for Sun's motion with respect to the local standard of rest, $(U,V,W)=(11.10 \pm 0.72, 12.24\pm0.47, 7.25\pm0.37)~\rm{km/s}$ \cite{Schonrich-2010}.

\subsection{Orbital Integration}\label{sec:methods:integration}
To infer the time evolution of the Milky Way satellite system, the orbits of the satellites are integrated numerically as massless test particles in a static potential using the Gala package \citep{gala}. The potential consists of a disk, stellar nucleus and bulge, and a dark matter halo. The disk is modelled as an
axisymmetric Miyamoto-Nagai disk
\citep{Miyamoto-Nagai-1975}, which, for our default model,
has disk mass $5.17\times 10^ {10}~\Ms$, a = 3 kpc, b = 0.028 kpc
\citep{Licquia-2015}. The nucleus and stellar bulge are both
modelled as spherically symmetric Hernquist profiles
\citep{Hernquist-1990}. For the nucleus we assume a mass of
$1.71 \times 10^9~\Ms$, and a scale radius $a = 0.07$~kpc, and for the
bulge we assume a mass of $5.0 \times 10^9 ~\Ms$ and $a = 1.0$
kpc. For the dark matter halo we assume a spherically symmetric
NFW \citep{Navarro-1996} potential.

Until recently, the Milky Way halo mass may have been a prohibitive
source of uncertainty for calculating the orbital evolution of the
satellites, as its value was known only to within a factor of two \citep[e.g.][]{Wang-2020}. However, the Galactic
halo mass has now been estimated with an uncertainty of only about 20\% using {\it Gaia} data. Multiple dynamical probes, such as the stellar rotation
curve, the escape velocity, the motions of halo stars, globular
clusters, and satellite galaxies
\citep{Monari-2018,Callingham-2019,deason-2019,Cautun-2020,Koppelman-2021}, consistently imply a dark matter halo mass for the MW of $M_{200}=(1.0\pm0.2)\times10^{12}~\Ms$ and NFW concentration, $c_{200}=11\pm2$.

Based on these results, we adopt a reference MW halo of mass $1.0 \times 10^{12}\Ms$ and a concentration parameter, $c_{200}=11$, corresponding to an NFW scale radius of $r_s=19.2$~kpc. The positions and velocities relative to the plane of satellites, and the orbital periods and apocentre distances for the default potential, are listed in Table~\ref{tab:kinematics}, where the quoted uncertainties reflect $68\%$ confidence intervals for all quantities based on Monte Carlo sampling. Varying the MW potential within the observational uncertainties does not significantly affect the conclusions of our study, as we show in Appendix~\ref{appendix:potential}.

\subsection{\LCDM simulations} \label{sec:methods:simulations}
The simulations used in this work are cosmological zoom-in constrained simulations, based on initial conditions created for the {\sc Sibelius} project \citep{Sawala-2021, McAlpine-2022} and designed to reproduce Local Group (LG) analogues within the observed large-scale structure. The simulations assume a \LCDM cosmology with $\Omega_0 = 0.307$, $\Omega_\Lambda = 0.693$, $\sigma_8 = 0.8288$, and $h = 0.6777$. We use physical units throughout this work. In total, we generated 60,000 simulations, resulting in several thousand loosely defined Local Group analogues. From these, we selected 101 for the re-simulations, performed at a mass resolution of $1.0\times10^6\Ms$ with the public {\sc Gadget-4} code \citep{Springel-2021}. At this resolution, a MW analogue halo contains approximately $10^6$ particles and an average of $\sim 200$ subhalos down to $2\times 10^7\Ms$ can be identified within 300 kpc from the centre.

Structures and self-bound substructures were identified using the Friends-of-Friends (FoF) and {\sc Subfind} algorithms implemented in {\sc Gadget-4} at 60 snapshots equally spaced in time, from $z=4$ until a lookback time of 1 Gyr, and a further 40 snapshots equally spaced over the final 1 Gyr up to $z=0$. Throughout this work, we refer to the two principal self-bound substructures of each LG analogue at $z=0$ simply as ``halos'', and to the lower mass substructures within 300 kpc of the centre of each halo as ``satellites''. For the purposes of this work, we consider both halos as Milky Way analogues.

We use {\sc Gadget}'s on-the-fly merger tree construction, and cut the chain of links when a subhalo's progenitor is no longer found, or when a clear discontinuity in mass and position indicates that a satellite's progenitor has been erroneously identified as the main halo. At each snapshot, we record the maximum circular velocity of each subhalo,
\vmax $= \mathrm{max}\left(\sqrt{ \frac{G M}{r}}\right) $, and define $v_\mathrm{peak}$ as the highest value of $v_\mathrm{max}$ of a subhalo and its progenitors over time. Following \cite{Libeskind-2005}, we use the standard procedure to rank satellites by $v_\mathrm{peak}$, and identify the top 11 within 300 kpc of each MW analogue at $z=0$ as analogues to the classical MW satellites.

\subsection{Orphan subhalos} \label{sec:methods:orphans}
As noted above, the radial distribution of satellites is important for the anisotropy. Numerical simulations suffer from the artificial disruption of substructures, that can affect subhalos far beyond the particle number limit at which they can theoretically be identified \citep{VanDenBosch-2018, Guo-2011}.

This effect can, however, be mitigated using semi-analytical models (which populate merger-trees constructed from simulated dark matter subhalos with galaxies). These models include so-called ``orphan" galaxies, that is, galaxies whose dark-matter subhalo has been numerically disrupted. After the subhalo is disrupted numerically, its subsequent evolution is followed by tracing the positions of its most bound particle \citep{Simha-2017}. Our ``complete" sample includes these ``orphan" subhalos.

One important result of this work is that the ``incomplete" and ``complete" samples of satellite halos have different radial distributions. Even though our high-resolution simulations resolve, on average, 200 surviving satellite halos inside 300 kpc of each MW analogue at $z=0$, and although we rank the satellites by \vpeak (\vmax being more strongly affected by tidal stripping), we find that the radial distribution of the top 11 surviving satellites in the ``incomplete" samples are systematically and significantly less centrally concentrated than the MW's brightest satellites.

\begin{figure}
    \centering
    \includegraphics[width=.8\columnwidth]{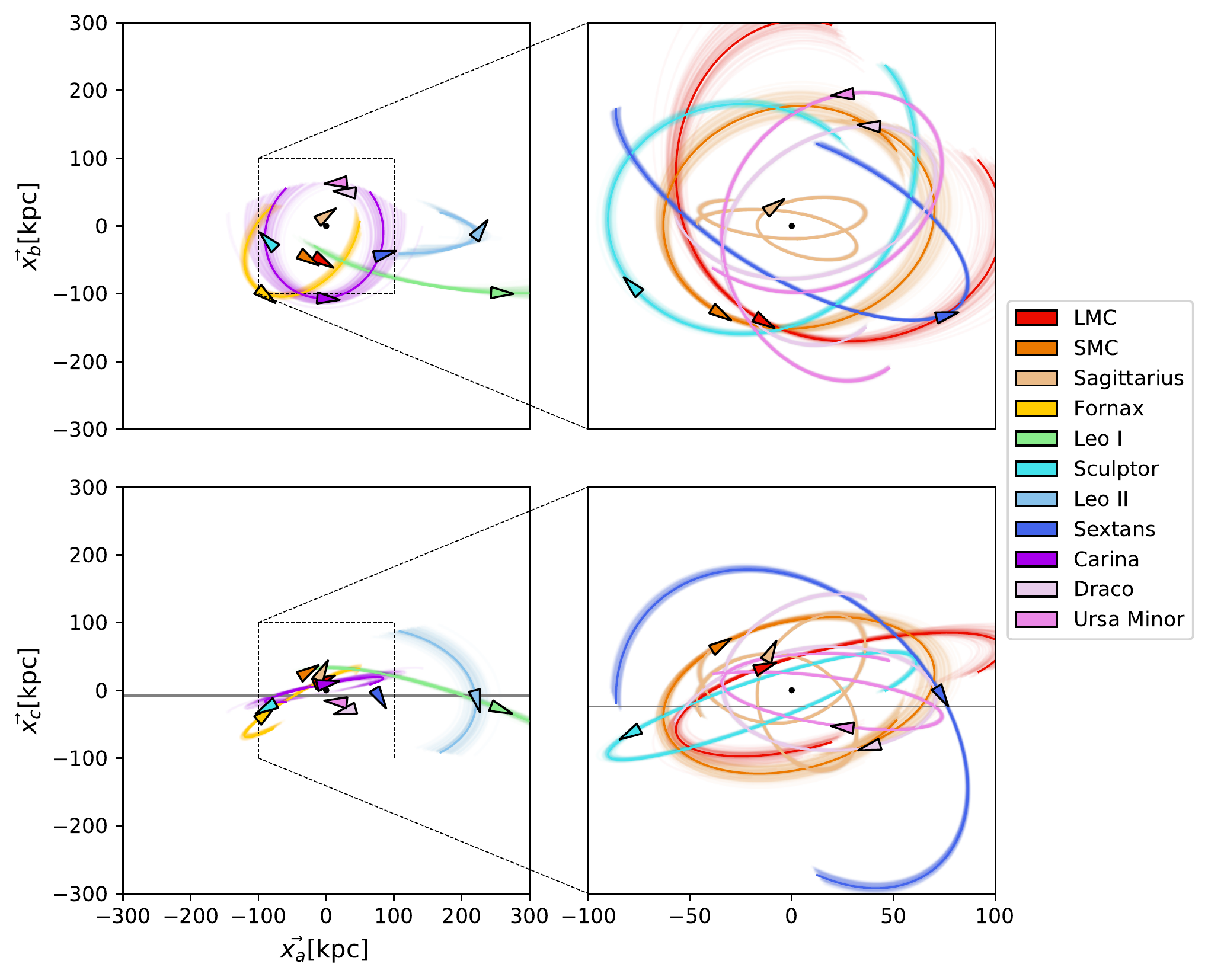}
    \caption{Maximum likelihood positions (arrowheads) and orbits of
      the 11 brightest MW satellites within 300 kpc, projected face-on
      (top) and edge-on (bottom) according to the eigenvectors of the inertia tensor. Bold lines show maximum-likelihood orbits integrated for 1~Gyr into
      the past and future in a halo of mass $10^{12}\Ms$, faint lines show 200 Monte-Carlo samples. The {\it Gaia} EDR3 measurements tightly constrain the proper motions, except for the LMC and SMC. Several galaxies, including the two outermost, Leo I and II, are presently crossing the plane (indicated by grey horizontal lines in the bottom panels), which soon disperses as a result.}
    \label{fig:projections}
\end{figure}

\section{The Milky Way's Plane of Satellites in light of {\it Gaia} EDR3} \label{sec:results}
Figure~\ref{fig:projections} shows the present most likely positions and estimated orbits of the 11 brightest MW satellites projected along the principal axes of inertia. For the present positions, we measure $c/a = 0.183 \pm 0.004$ and $(c/a)_\mathrm{red}=0.3676 \pm 0.0004$. In the bottom two panels, the solid grey line shows the plane of satellites projected edge-on. However, from a visual inspection of the orbits, shown here integrated over $\pm$ 1 Gyr and including Monte-Carlo sampling of the observational uncertainties, it is already apparent that this configuration is short lived.

\begin{figure}
\centering
    \includegraphics[width=\columnwidth]{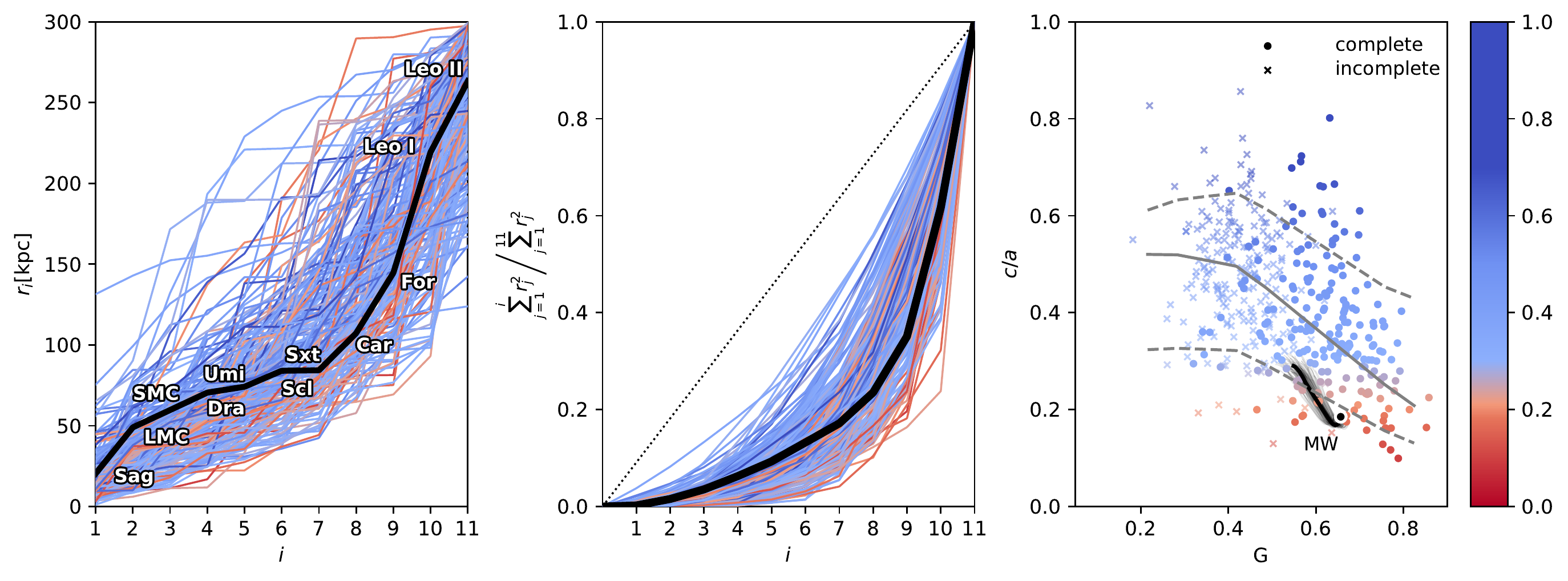}
    
    \caption{Radial distribution and anisotropy of the classical MW satellites and those of simulated \LCDM counterparts. On all panels, black symbols and lines represent the MW, lines coloured by $c/a$ represent the simulations. Left panel: radius, $r_i$, of the $i^{\mathrm{th}}$ closest satellite. Centre panel: sum of the squares of the radii of the closest $i$ satellites, normalised by the sum of all 11 satellites, i.e. the cumulative contributions to the inertia. The {\it Gini coefficient of inertia}, $G$, corresponds to the area between each line and the diagonal. Right panel: correlation between $G$ and anisotropy, $c/a$, accounting for artificial disruption (``complete'', circles), or without accounting for artificial disruption (``incomplete'', crosses). Grey lines indicate median and 10$^\mathrm{th}$ and 90$^\mathrm{th}$ percentiles. The black circle denotes the Milky Way's present values of $G$ and $c/a$, lines show its most likely (bold) and Monte-Carlo sampled (thin) evolution over the past 0.5 Gyr. Accounting for artificial disruption, the MW lies within the distribution.}
    \label{fig:radial}
\end{figure}

\subsection{Spatial Anisotropy} \label{sec:spatial}
Earlier comparisons to \LCDM systems \citep{Pawlowski-2019a} found that only $0.7\%$ of \LCDM simulations produce systems as anisotropic as the Milky Way. However, we find this to be an artefact caused by the disruption of satellites in numerical simulations, which results in artificially extended radial profiles \citep{Guo-2014, VanDenBosch-2018, Webb-2020}. Accounting for this effect through the inclusion of orphans (see Section~\ref{sec:methods:orphans}) into our ``complete" sample of satellites, we recover radial distributions resembling the MW's, as shown in the left panel of Figure~\ref{fig:radial}.

The right panel of Figure~\ref{fig:radial} shows the relationship between $G$ and $c/a$. Systems with higher central concentration (higher $G$) tend to be more anisotropic (lower $c/a$). Accounting for artificial disruption (filled circles), $58\%$ of \LCDM systems have $G$ above the MW, and 11 ($5.5\%$) have $c/a < 0.183$. Neglecting this effect (faint crosses) produces no systems with $G$ as high the MW and only two (1\%) with as low $c/a$ - in line with earlier studies  \citep{Shao-2019, Pawlowski-2019a} which found $c/a$ values as low as the MW's to be exceedingly rare.

\begin{figure}
    \includegraphics[width=\columnwidth]{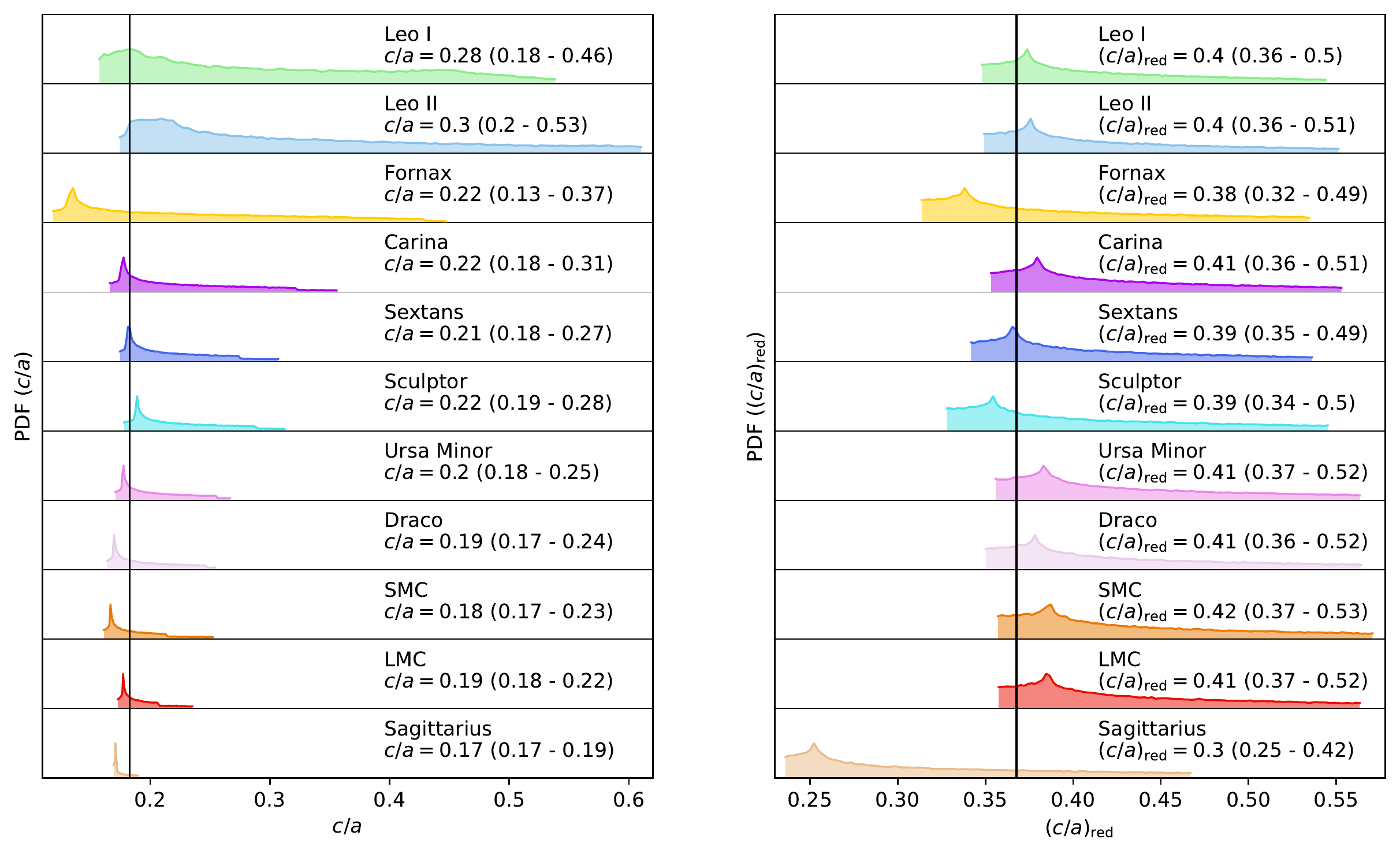}
    \caption{Probability density functions of $c/a$ (left) and
    $(c/a)_\mathrm{red}$ (right) for the 11 brightest satellites when
    the angular coordinates of each galaxy are randomised in turn,
    with the distance set to the observed value, and the coordinates
    of all other galaxies kept fixed. Numbers show the median values of $c/a$, and
    $(c/a)_\mathrm{red}$; those in brackets show the 
    $10^\mathrm{th}$ and $90^\mathrm{th}$ percentiles. Galaxies are
    sorted from top to bottom in decreasing order of radius. Black
    horizontal lines indicate the vertical offset, the black vertical
    lines show the values with all 11 galaxies at their observed
    positions. Each galaxy impacts the distribution of $c/a $
    differently, and the range of possible $c/a$ values correlates
    with the radius of the satellite. Just placing either one of Leo I
    or Leo II at 
    different angular coordinates at their respective radius could
    result in a completely different anisotropy, including cases that
    are more
    isotropic than the majority of \LCDM systems. }
    \label{fig:anisotropy-distributions}
\end{figure}

The outsized influence of the outermost satellites on the measured anisotropy is shown in Figure~\ref{fig:anisotropy-distributions}, which shows the probability distributions of $c/a$ (left panel) and $(c/a)_\mathrm{red}$ (right panel) when one satellite is placed at random angular coordinates at its observed radius, while all other satellites remain fixed. Satellites are ordered from top to bottom in order of decreasing distance. Vertical lines show the values for all 11 satellites at their observed positions. We also list the median values of the distributions and, in brackets, the range corresponding to $1\sigma$ around the median. For Sagittarius, the satellite with the smallest distance, the $c/a$ distribution is extremely narrow: due to its close proximity, it contributes less than $1\%$ to the inertia tensor. For Fornax, the third most distant galaxy, located 38 kpc above the plane, randomising the angular coordinates can result in both significantly greater or smaller anisotropy. However, most significantly, for the two most distant satellites, Leo I and Leo II, randomising the angular coordinates of just one object raises the median value of $c/a$ to 0.28 and 0.31, respectively, with maxima of 0.53 and 0.63. In other words, randomising the position of Leo I or Leo II alone could turn the Milky Way's classical satellites into a system more {\it isotropic} than the majority of \LCDM systems.

In addition to the radial distribution, the Milky Way's present anisotropy results from the fact that its two outermost satellites, Leo I and Leo II, which contribute two thirds of the total inertia, are currently in close proximity to each other. However, as is already apparent from Figure~\ref{fig:projections}, and as we discuss in more detail below, this constellation is short-lived.

\clearpage

\begin{figure}
    \includegraphics[width=\columnwidth]{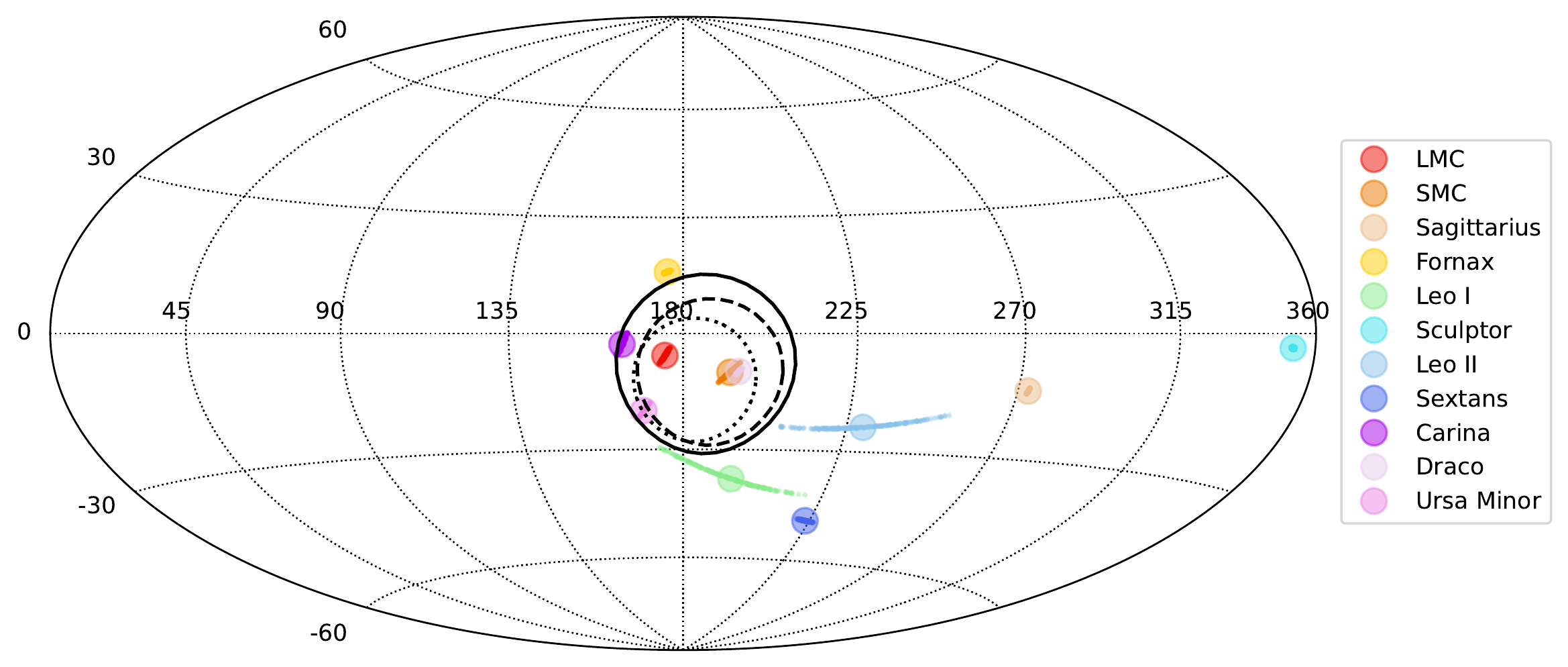}
    \caption{Hammer projection of Milky Way satellite orbital poles, using {\it Gaia} EDR3 proper motions. Large circles show the most likely values, small circles show 200 Monte Carlo samples of the observational errors, within $\pm 1 \sigma$ of the most likely values. The dotted black curve indicates the dispersion reported by \cite{Pawlowski-2019a} for seven MW satellites. The solid black curve indicates the dispersion we find for the same set based on {\it Gaia} EDR3, the dashed black curve indicates the minimum dispersion we find for seven satellites exchanging Leo I and Leo II. The orbital poles of the MW satellites are significantly clustered, but several of our simulated \LCDM systems contain equally or more strongly clustered satellite systems.}
    \label{fig:hammer}
\end{figure}

\subsection{Orbital Anisotropy} \label{sec:orbital}
Supporting the notion that the satellite plane constitutes a spinning disk, the orbital poles of 7 of the 11 classical satellites -- the LMC, the SMC, Fornax, Leo II, Carina, Draco and Ursa Minor --  are reportedly clustered with a standard deviation in direction of only $\Delta_{\rm{std}} (N_s=7)=16.0^{\circ}$, found in only $0.04\%$ of \LCDM systems \citep{Pawlowski-2019a}.

Using the more precise proper motions from {\it Gaia} EDR3 \citep{McConnachie-2020b} for the same seven satellites, we find that this angle increases significantly, to $\Delta_{\rm{std}} (N_s=7)=23.2^{\circ}~_{-2.8}^{+3.5}$. This configuration has a 0.087\% probability to arise from an isotropic distribution. We also repeated the analysis, and find that a different subset (that includes Leo I instead of Leo II) yields a smaller dispersion of $\Delta_{\rm{std}} (N_s=7)=18.9^{\circ}~_{-1.4}^{+1.9}$, with a corresponding 0.011\% probability. In Figure~\ref{fig:hammer}, we show the orbital poles of the 11 classical satellites, and the orbital pole dispersions calculated by \cite{Pawlowski-2019a} (dotted), by us for the same set of seven satellites based on {\it Gaia} EDR3 (solid), and by us for the most clustered set of seven satellites (dashed).

In our sample of 202 simulated systems, Among our sample of 202 simulated systems, adopting either  $\Delta_{\rm{std}} (N_s=7) = 18.9^{\circ}$ or $23.2^{\circ}$ and accounting for the minimum look-elsewhere effect (see Section~\ref{sec:methods:look-elsewhere}), we find three or five systems with subsets of satellites with a smaller probability to arise from isotropic distributions. That is, we find that $\sim 2\%$ of \LCDM systems contain satellites whose orbital poles are even more anisotropic than the most clustered subset of the Milky Way, a $\sim$ 50-fold increase over previous results. The orbital clustering of a subset of the Milky Way satellites is unusual in \LCDM, but not {\it astronomically} improbable.

Importantly, while the ``plane of satellites" includes all 11 classical satellites, the orbital anisotropy only concerns a subset, which is in fact more spatially isotropic than the system as a whole. The orbital pole clustering does not drive the spatial anisotropy.

Importantly, while the ``plane of satellites" includes all 11 classical satellites, the orbital anisotropy only concerns a subset, which happens to be more spatially isotropic than the system as a whole. The orbital pole clustering does not drive the spatial anisotropy.

\subsection{Time Evolution} \label{sec:evolution}
Another defining feature of a rotationally supported disk would be a significantly higher velocity dispersion parallel to the plane than perpendicular to it. However, for the MW's classical satellites, we measure $\sigma_{v\parallel} = 165.1 \pm 1.2$~kms$^{-1}$ and $\sigma_{v\perp} = 121.6 \pm 0.4 $~kms$^{-1}$. The ratio, $\sigma_{v\parallel} / \sigma_{v\perp} = 1.36$, is indistinguishable from the purely geometrical factor of $\sqrt{2}$. By this basic measure, the plane is not rotation-supported.

The longevity of the plane can also be tested directly via orbital integration, as described in Section~\ref{sec:methods:integration}. This method was first, but inconclusively, applied using pre-{\it Gaia} data by \cite{Maji-2017}. In this work, as described in Section~\ref{sec:methods:observations}, we benefit from significantly more precise observations, including {\it Gaia} EDR3 proper motions and more accurate distances.

\begin{figure}
    \centering
    \includegraphics[width=\textwidth]{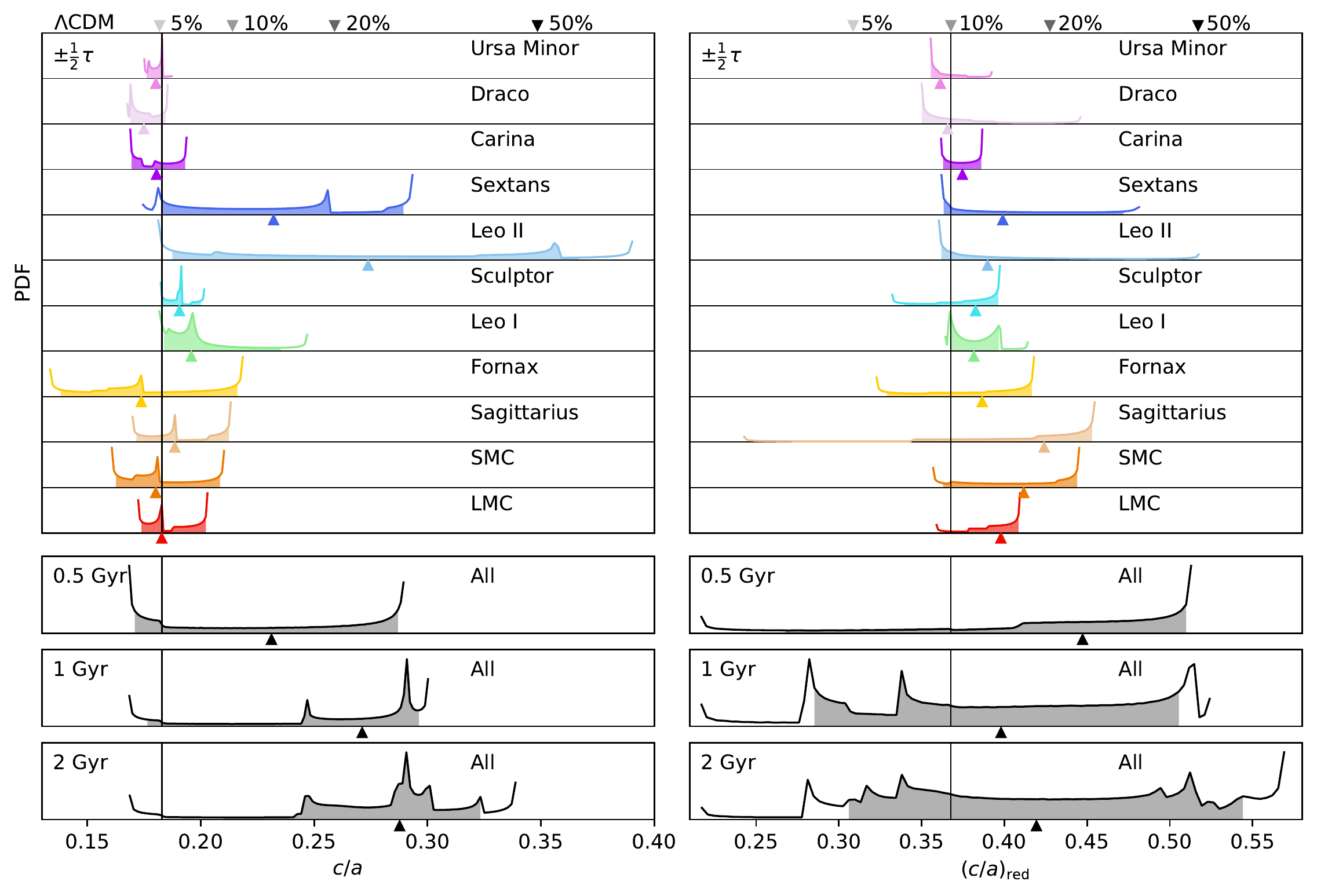}
    \caption{Top: time-averaged probability densities for the
      anisotropy, $c/a$ (left) and $(c/a)_\mathrm{red}$ (right), when each satellite evolves along its most likely orbit during one period, $\tau$, while the other 10
      satellites remain at their present positions. Bottom: probability density functions over lookback times of 0.5, 1 and 2 Gyr, evolving all orbits
      simultaneously. Triangles below each graph indicate the time-averaged medians, filled areas extend from the 10$^\mathrm{th}$ to 90$^\mathrm{th}$ percentiles. Black vertical lines indicate the present anisotropy, $c/a=0.183$, $(c/a)_\mathrm{red}=0.364$. Downward triangles at the top show percentiles in the \LCDM simulations. Over the past 1~Gyr, $c/a$ varied between 0.17 and 0.31, while $(c/a)_\mathrm{red}$ varied between 0.22 and 0.52. The present value of $c/a$ is an outlier even compared to the past 0.5 Gyr.}
    \label{fig:distributions}
\end{figure}

In Figure~\ref{fig:projections}, we saw that several satellites are presently crossing the plane, while Leo I and II, which dominate the inertia, are moving apart. To elucidate the impact of such ``fortuitous alignments" on the anisotropy, we show at the top of Figure~\ref{fig:distributions} the anisotropy distributions when each satellite moves along its orbit while the others remain at their present positions. The time-averaged anisotropy of the system is then calculated over one full orbital period centred on the present time. Depending on the orbital phase of Leo II alone, $c/a$ could be as high as 0.39, more {\it isotropic} than most \LCDM systems.

The bottom panels of Figure~\ref{fig:distributions} show time-averaged probability densities of $c/a$ and $(c/a)_\mathrm{red}$ when all satellites evolve simultaneously. The current value of $c/a$ is significantly lower than in the recent past: over the past 0.5 and 1~Gyr, the time-averaged medians of $c/a$ are 0.23 and 0.27 respectively, greater than $13\%$ and $23\%$ of \LCDM systems. $(c/a)_\mathrm{red}$ has varied widely. Neither metric is an invariant of the satellite system. Instead, both are sensitive to the orbital phases.

\begin{figure}[h!]
    \includegraphics[width=\columnwidth]{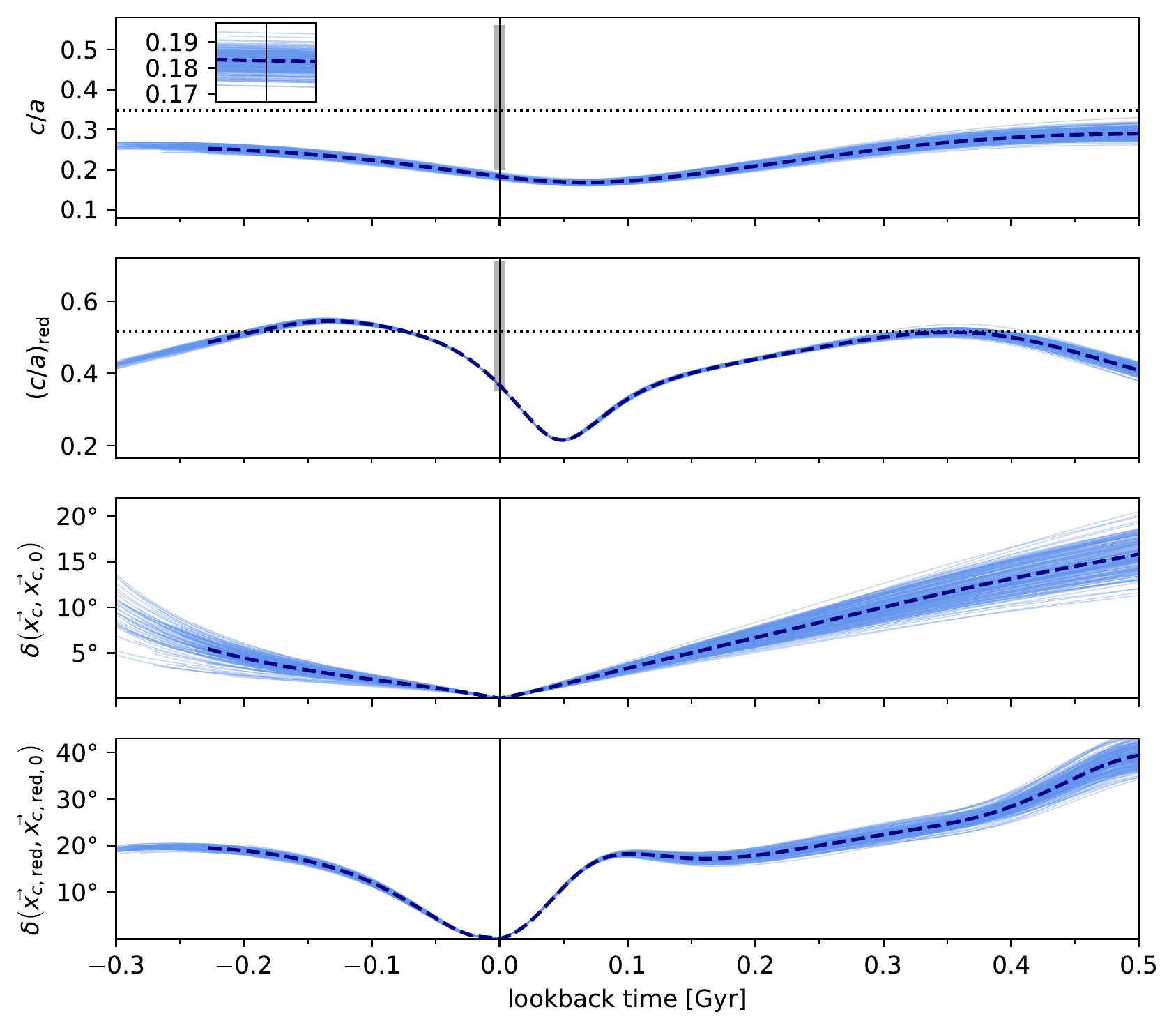}
    \caption{Evolution of $c/a$, $(c/a)_{\mathrm{red}}$, and directions of the normals to the plane of the full and reduced inertia tensors $\vec{x_c}$ and $\vec{x_{c,\mathrm{red}}}$. Dashed dark blue lines show most likely results, light blue lines show 200 Monte-Carlo samples, until one satellite is beyond 300 kpc from the Galactic centre. The small inset shows $t=0 \pm 1$~Myr. Grey vertical bars at $t=0$ show 10$^{\mathrm{th}}$ to 90$^{\mathrm{th}}$ percentiles of simulated \LCDM analogues at $z=0$, dotted horizontal lines show the medians of these distributions. The MW's $c/a$ evolves towards the median of the \LCDM systems, while $(c/a)_{\mathrm{red}}$ varies significantly, exceeding the median in both the near past and near future. The evolution of $\vec{x_c}$ and $\vec{x_{c,\mathrm{red}}}$ reveal the plane to be tilting by either definition.}
    \label{fig:MC-evolution}
\end{figure}

The four panels of Figure~\ref{fig:MC-evolution} show the evolution of $c/a$, $(c/a)_\mathrm{red}$, and of the orientations of the planes defined by the full and reduced inertia tensors, which we parametrise by the angles between the vectors normal to the planes, $\vec{x_c}$ and $\vec{x_{c,\mathrm{red}}}$ and their present day equivalents, $\vec{x_{c,0}}$ and $\vec{x_{c,0,\mathrm{red}}}$. The value of $c/a$ approaches the \LCDM median within a lookback time of 0.5 Gyr. $(c/a)_\mathrm{red}$ evolves more rapidly, twice exceeding the \LCDM median. That $c/a$ and $(c/a)_\mathrm{red}$ vary on such different timescales is a further consequence of the radial distribution. While Leo I and Leo II, which largely determine $c/a$, have orbital periods of $3.4 \pm 0.2$ and $7.2 \pm 0.3$ Gyr, respectively, the eight closest satellites, which largely determine $(c/a)_\mathrm{red}$ have orbital periods under 2~Gyr.

\begin{figure}[h!]
    \includegraphics[width=\columnwidth]{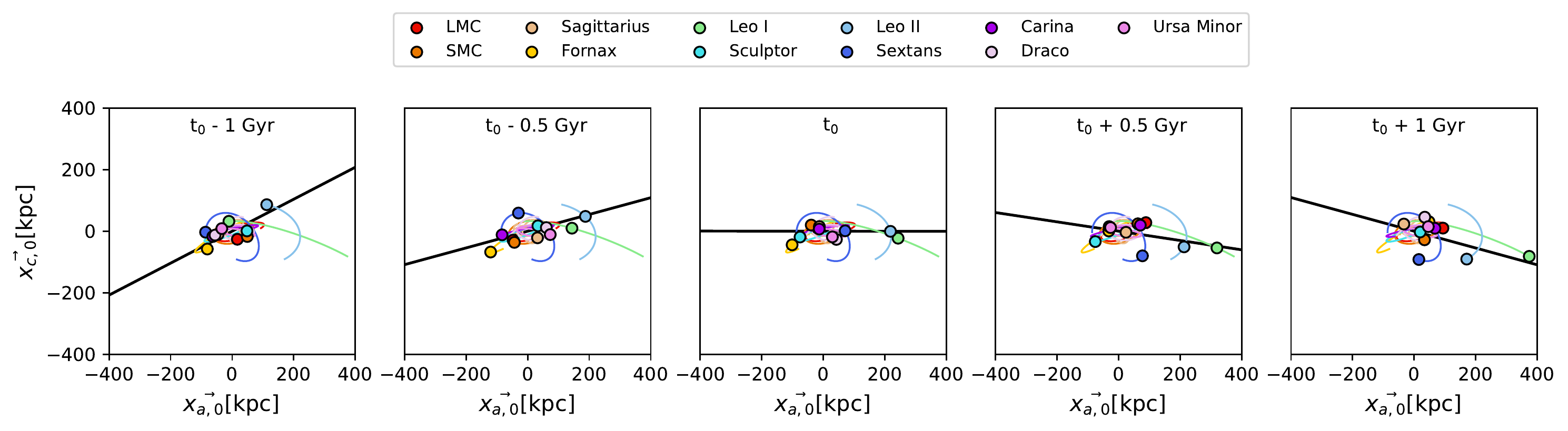}
    \caption{Evolution of the orientation of the plane of satellites. Shown on all panels are the most likely orbits of the MW satellites for $\pm$ 1 Gyr from the present time, analogous to Figure~\ref{fig:projections}. From left to right, panels show the position of satellites at 1 Gyr and 0.5 Gyr in the past, today (denoted with $t_0$), and 0.5 and 1 Gyr into the future. All panels are plotted in the frame of the major and minor eigenvectors of the inertia tensor at time $t_0$, $\vec{x_{a,0}}$ and $\vec{x_{c,0}}$. Thick black lines show the plane at each time edge-on. In the centre panel, by construction the plane axes align with the coordinate axes. In the other panels, the tilt of the plane can be observed. The orientation of the plane closely follows the locations of the most distant satellites.}
    \label{fig:plane-tilt}
\end{figure}

We further see that the orientation of satellite plane is not stable, but has tilted by $\sim 17^{\circ}$ over the past 0.5 Gyr (and $\sim 40^{\circ}$ for the reduced definition). The tilt of the plane is illustrated most clearly in Figure~\ref{fig:plane-tilt}, where we show the most likely orbits and positions of the classical MW satellites, projected in the frame of the major and minor eigenvectors at time $t_0$. In each panel, the thick black line shows the orientation of the plane edge-on, computed at the respective time. In the centre panel, by construction the plane axes align with the coordinate axes. In the other panels, the tilt of the plane can be observed. Rather than satellites orbiting inside a stable plane, the plane itself tilts, as it tracks the positions of its two most distant members. Neither observational errors (see Appendix~\ref{appendix:gaia}) nor uncertainties regarding the MW potential (see Appendix~\ref{appendix:potential}) significantly affect these results.

\section{Summary} \label{sec:summary}
The high reported anisotropy of the MW satellite system can largely be attributed to its high central concentration, not previously reproduced in simulations, combined with the close but fleeting contiguity of its two most distant members. Accounting for the radial distribution reveals the MW satellites to be consistent with \LCDM expectations. Compared to previous works, we also find a much higher likelihood of subsets whose orbital poles are as clustered as the MW. Although the Milky Way contains such a subset, the plane of satellites does not constitute a rotationally supported disk. Instead, it evolves on timescales similar to the ``transient" alignments previously found in \LCDM simulations.

Our orbital integration assumes a static MW potential with satellites as massless test particles. We Monte-Carlo sample all sources of observational error and also vary the components of the potential within the uncertainties (see Section~\ref{sec:methods:observations} and Appendix~\ref{appendix:gaia}). We find our results to be robust, and while the real potential is more complex, for example due to the presence of the LMC, these simplifications are valid within a dynamical time ($\sim 2$ Gyr) of the halo, particularly for the important outer satellites \citep{Garavito-Camargo-2021, Battaglia-2022}. A more complex potential would only accelerate the dissolution of a rotating disk \citep{Nuwanthika-2018}. The true Milky Way potential also evolves with time, but the dynamical time of the halo ($\sim 2$ Gyr at $z=0$) is significantly longer than the time scale for the reported dissolution of the plane of satellites (several hundred Myr).

This work only directly addresses the archetypal ``plane of satellites" around the MW. Anisotropic satellite distributions have also been reported around several other galaxies \citep{Ibata-2014, Muller-2021} with different criteria, which can exacerbate the look-elsewhere effect. Assessing their significance requires a careful statistical analysis \citep{Cautun-2015}. While not all criteria are equally sensitive to the radial distribution, we also expect that the significantly higher anisotropy we report here for simulated MW analogues will apply to \LCDM analogues of other, similarly defined systems.

After centuries of observations, the Milky Way and its satellites are the best studied system of galaxies in the Universe. Viewed in sufficient detail, every system inevitably reveals some remarkable features, and the Milky Way is no exception. However, based on the best currently available data, there is no evidence for a plane of satellites incompatible with, or even particularly remarkable in \LCDM. On the contrary, as judged by the spatial anisotropy of the brightest satellite galaxies, we live in a fairly typical \LCDM system.

\subsection*{Acknowledgements}
This work used the DiRAC Data Centric system at
Durham University, operated by the Institute for Computational Cosmology on behalf of the STFC DiRAC HPC Facility (www.dirac.ac.uk), and facilities hosted by the CSC - IT Centre for Science, Finland. The DiRAC system was funded by BIS National E-infrastructure capital grant ST/K00042X/1, STFC capital grants ST/H008519/1 and ST/K00087X/1,
STFC DiRAC Operations grant ST/K003267/1 and Durham University. DiRAC is part of the National E-Infrastructure.

TS is an Academy of Finland Research Fellow. This work was supported by Academy of Finland grant numbers 314238 and 335607. PHJ acknowledges the support by the European Research Council via ERC Consolidator Grant KETJU (no. 818930) and the Academy of Finland grant 339127. MS is supported by the Netherlands Organisation for Scientific
Research (NWO) through VENI grant 639.041.749. CSF acknowledges
support by the European Research Council(ERC) through Advanced
Investigator DMIDAS (GA 786910). GL acknowledges financial support by
the French National Research Agency for the project BIG4, under
reference ANR-16-CE23-0002, and MMUniverse, under reference
ANR-19-CE31-0020.

\subsection*{Data and Code}
The analysis in this paper was performed in python3 and makes extensive use of open-source libraries, including Matplotlib 3.4.2, NumPy 1.21.1 \citep{numpy}, SciPy 1.7.0 \citep{2020SciPy-NMeth}, GalPy 1.7.0 \citep{galpy}, Py-SPHViewer \citep{pysphviewer}, TensorFlow \citep{tensorflow2015-whitepaper} and Gala 1.4.1 \citep{gala}.
The full data and analysis code is available at   \url{https://github.com/TillSawala/plane-of-satellites}.

\bibliography{paper}

\clearpage

\section*{Appendices}

\setcounter{figure}{0}
\renewcommand\thefigure{A\arabic{figure}}
\renewcommand\thetable{A\arabic{table}}

\appendix

\section{Observational data and derived kinematics}\label{appendix:tables}
The observational data used in this paper, described in Section~\ref{sec:methods:observations} are listed in Table~\ref{tab:observations} together with their references. The positions and velocities relative to the plane of satellites, and the orbital periods and apocentre distances for the default potential, are listed in Table~\ref{tab:kinematics}, where the quoted uncertainties reflect $68\%$ confidence intervals for all quantities based on Monte Carlo sampling the observations.

\section{Different sources of proper motions}\label{appendix:gaia}
\begin{figure*}
    \includegraphics[width=3.in]{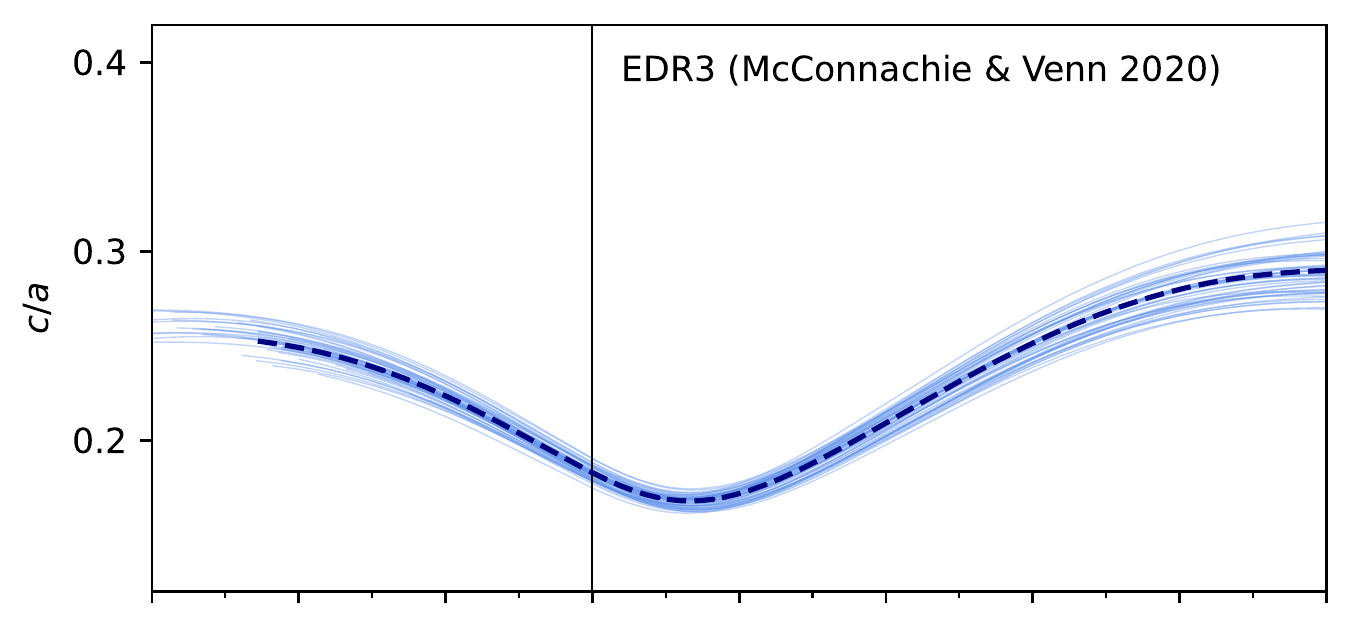}
    \includegraphics[width=3.in]{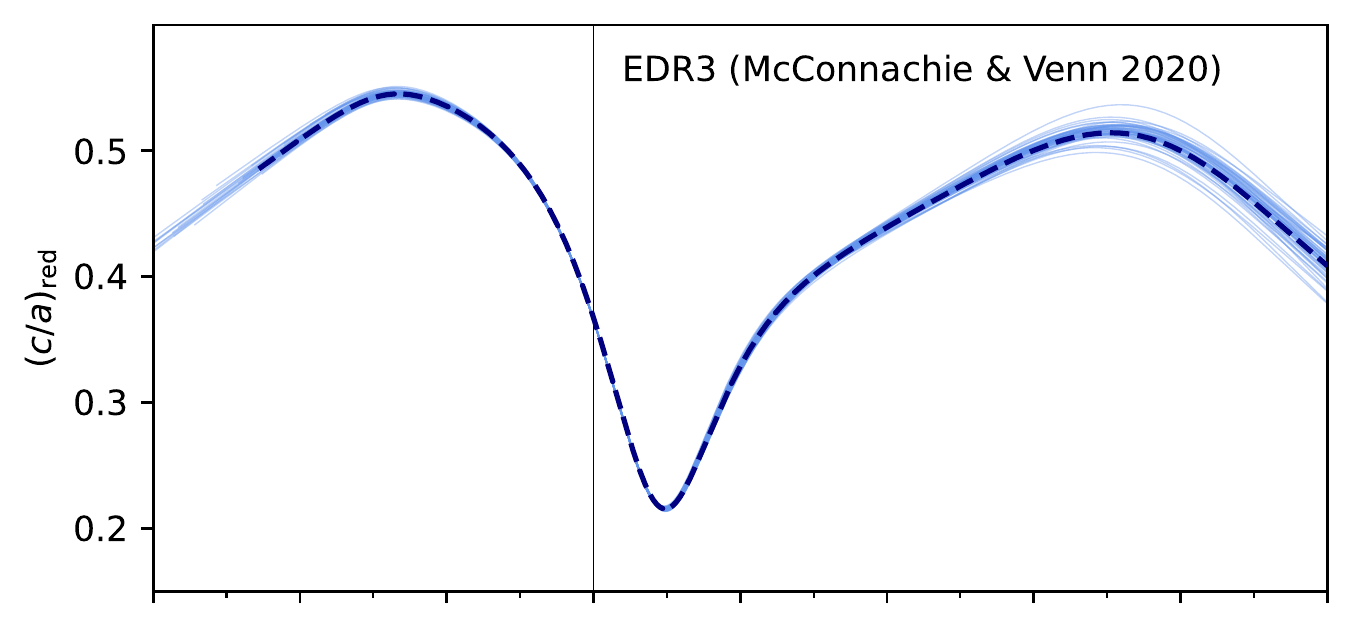}\\
    \includegraphics[width=3.in]{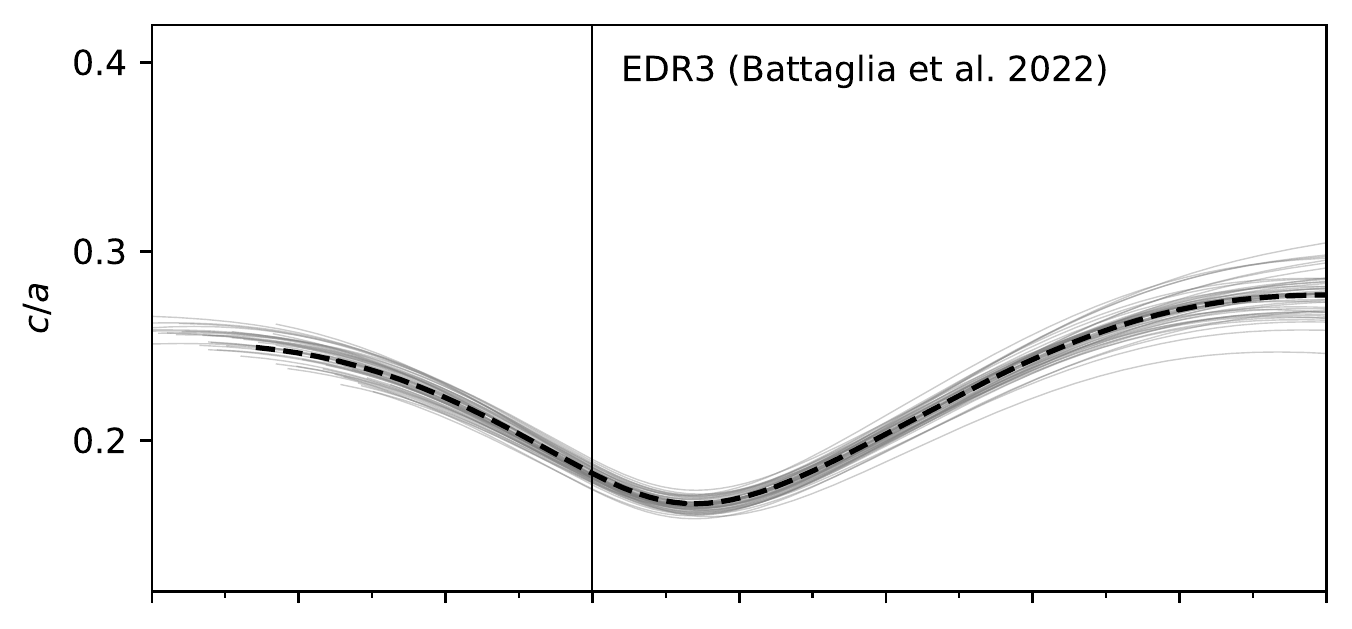}
    \includegraphics[width=3.in]{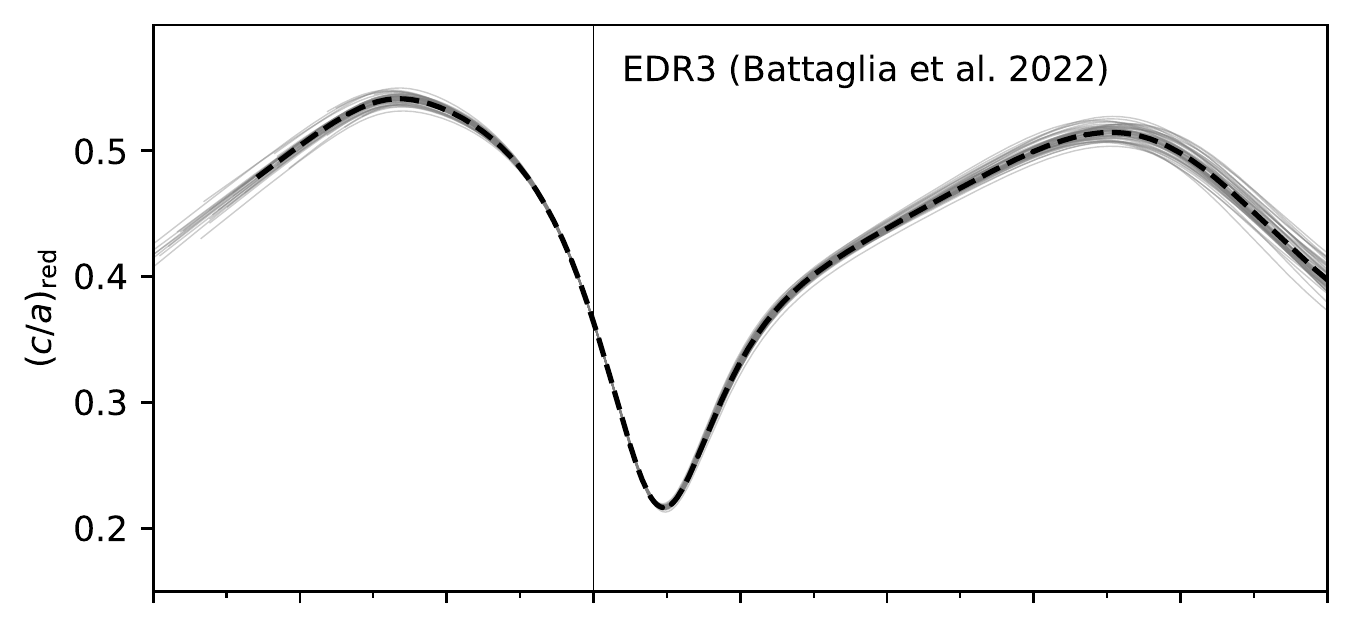} \\
    \includegraphics[width=3.07in]{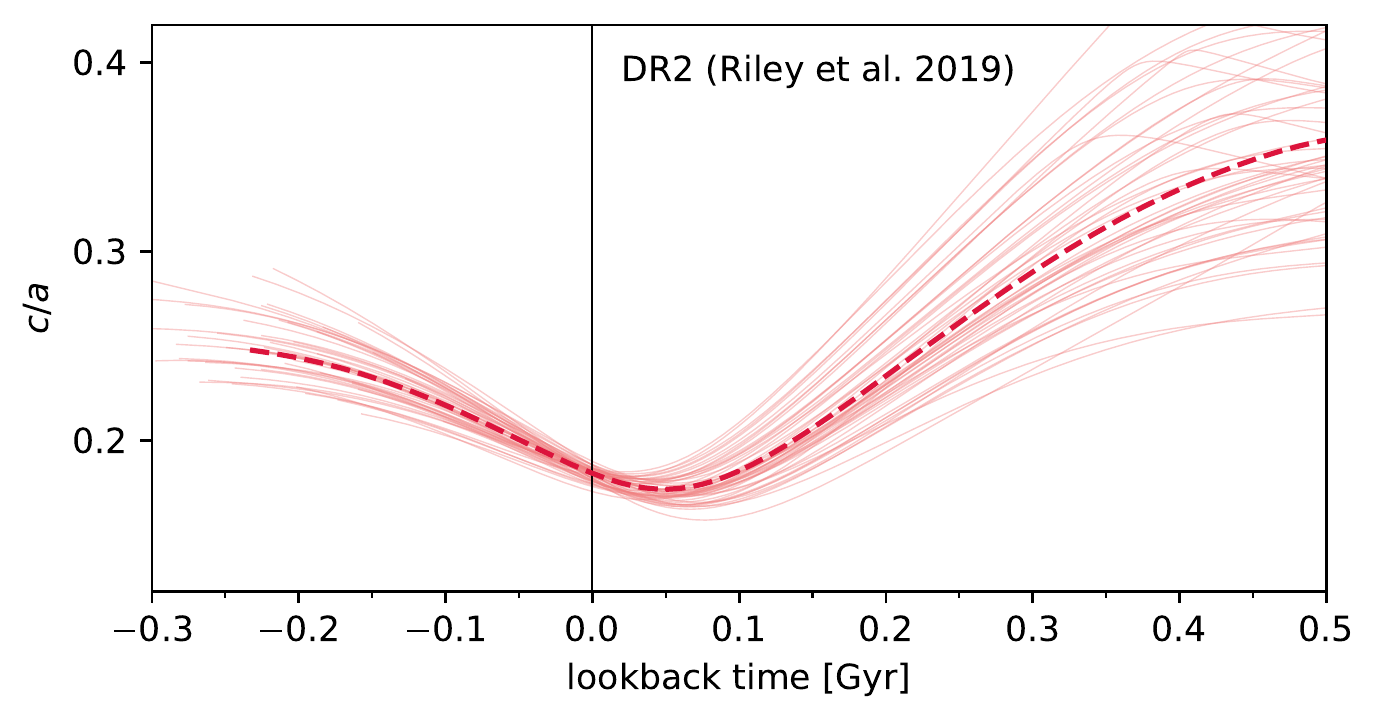} \hspace{-1.3mm}
    \includegraphics[width=3.1in]{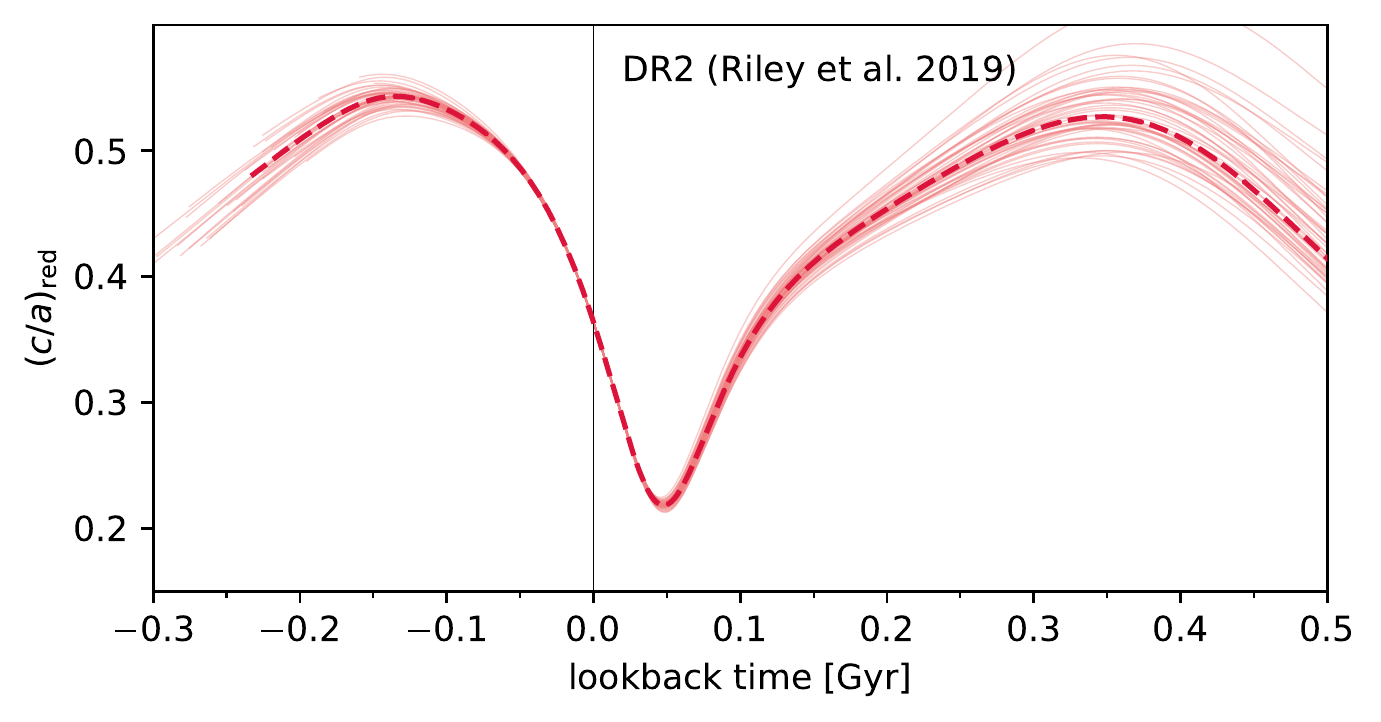}\\
    
    \caption{Evolution of $c/a$ (left column) and $(c/a)_{\mathrm{red}}$ (right column), from integrated satellite orbits, analogous to Figure~\ref{fig:MC-evolution}, showing the effect of using {\it Gaia} EDR3 proper motions by \cite{McConnachie-2020b} (top, blue lines), {\it Gaia} EDR3 proper motions by \cite{Battaglia-2022} (middle, black lines), or {\it Gaia} DR2 proper motions as described in \cite{Riley-2019} (bottom, red lines). For all data sets, thick dashed lines are the for the most likely observations, thin lines show 50 Monte Carlo samples. Lines extend to 0.5 Gyr into the past, and as long as all 11 satellites remain within 300 kpc of the centre into the future. The two EDR3 data sets are in excellent agreement. The main difference in the DR2 data is the larger errors, but the evolution of both $c/a$ and  $(c/a)_{\mathrm{red}}$ is essentially the same in all three data sets.}
    \label{fig:evolution-dr2-dr3}
\end{figure*}

We repeated our analysis using the {\it Gaia} EDR3 proper motions of \cite{Battaglia-2022} and the {\it Gaia} DR2 proper motions described in \cite{Riley-2019}. A comparison of the evolution of $c/a$ and $(c/a)_{\mathrm{red}}$, analogous to Figure~\ref{fig:MC-evolution}, is shown in Figure~\ref{fig:evolution-dr2-dr3}. Unsurprisingly, the evolution based on the two EDR3 data sets are in excellent agreement. The main difference when using the DR2 data is the larger uncertainty (compared to DR2, the astrometry errors of EDR3 are reduced by approximately a factor of two), but the evolution of both $c/a$ and $(c/a)_{\mathrm{red}}$ is essentially the same in all three data sets, and the results are consistent within the respective errors.

\section{The effect of uncertainty in the potential}\label{appendix:potential}
\begin{figure*}
    \includegraphics[width=\columnwidth]{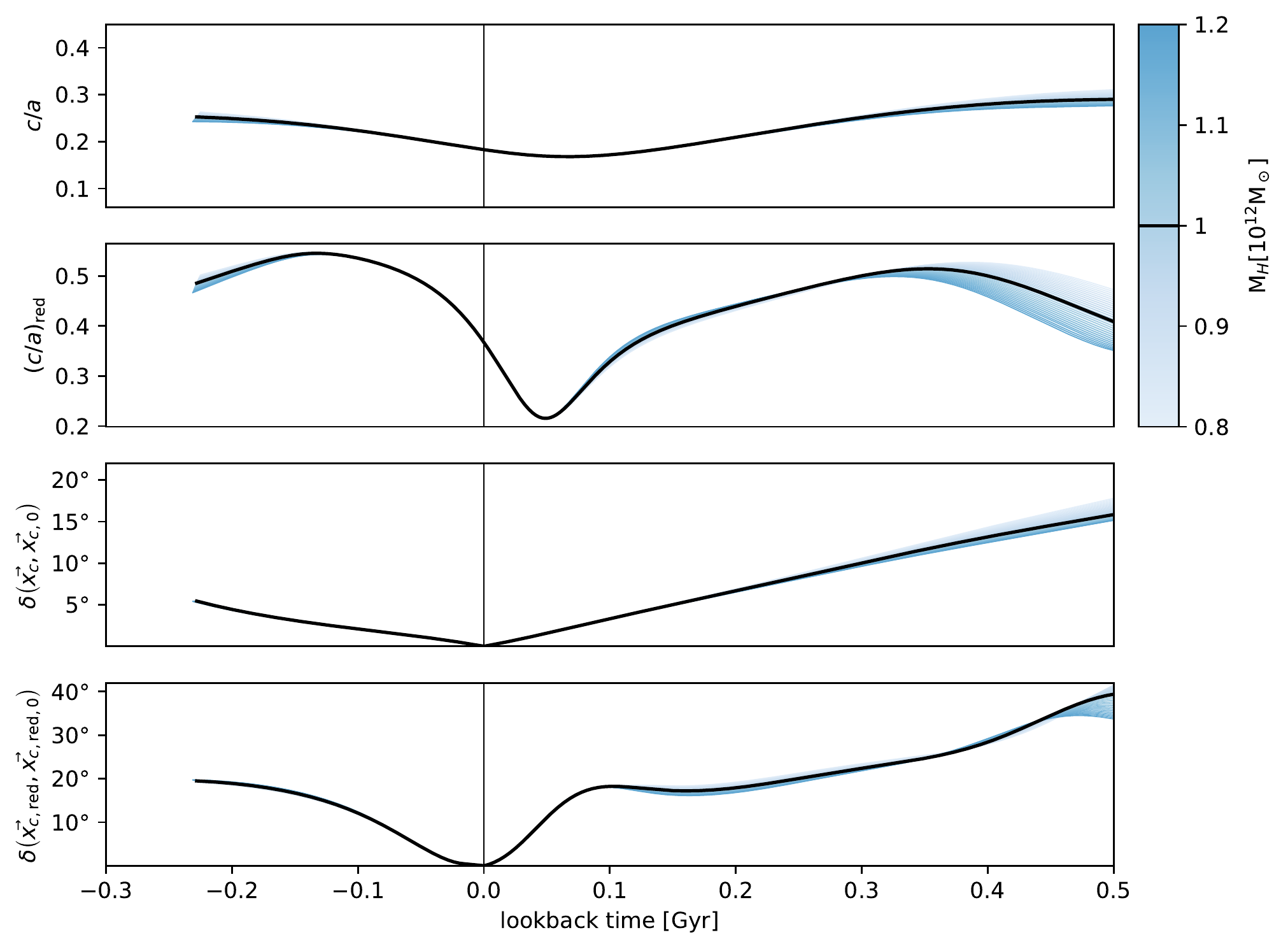}
    \caption{Evolution of $c/a$, $(c/a)_{\mathrm{red}}$ and the
    direction of the vectors normal to the planes of the full and
    reduced inertia tensors, from integrated satellite orbits,
    analogous to Figure~\ref{fig:MC-evolution} of the main part. Here,
    all lines are for the most likely observations, and we show the
    results of varying the mass of the dark matter halo. Black lines
    are for the default mass of $10^{12}\Ms$, shades of blue show the
    result of varying the mass between $0.8 - 1.2 \times
    10^{12}\Ms$. Lines extend to 0.5 Gyr into the past, and as long as
    all 11 satellites remain within 300 kpc of the centre into the
    future. The ratio $c/a$ reverts towards the median of the \LCDM values and $(c/a)_{\mathrm{red}}$ evolves rapidly for any mass. The tilt of the plane defined by either the full or reduced inertia tensor is essentially independent of the halo mass.}
    \label{fig:evolution-halomass}
\end{figure*}

In Figure~\ref{fig:evolution-halomass} we show the effect of varying the halo mass between $0.8$ and $1.2 \times 10^{12}~\Ms$. The inferred anisotropies, as measured by $c/a$ and $(c/a)_\mathrm{red}$, begin to diverge for lookback times beyond $\sim500$~Myr, but the behaviour is qualitatively similar across the halo mass range. We also tested the effect of varying the other three mass components (galaxy disk, bulge and nucleus) and the concentration parameter within the observational uncertainties, and found no significant effects. 
\clearpage

\begin{table}
\centering
\caption{Observed properties and their uncertainties for the Milky
Way's 11 brightest satellites, ordered by V-band magnitude, $M_V$. 
$\mu$ is the distance modulus, RA and Dec are coordinates of right
ascension and declination in degrees, pm$_\mathrm{RA}$ and
pm$_\mathrm{Dec}$ are the proper motions in RA and Dec in units of mas
yr$^{-1}$. The references are [1] \cite{McConnachie-2012} for the sky
coordinates, [2] \cite{Riley-2019} and [3] \cite{McConnachie-2020b} for the proper motions, and [4] \cite{Pietrzynski-2019}, [5] \cite{Graczyk-2020}, [6] \cite{Hernitschek-2019} [7] \cite{Freedman-2020}, [8] \cite{Martinez-Vazques-2015} and [9] \cite{Gullieuszik-2008} for the distance moduli. } 
\label{tab:observations}
\vspace{.5cm}

\resizebox{\columnwidth}{!}{%
\begin{tabular}{llllllll}
\hline\hline   
Name & $M_V$ & $\mu$ & RA & Dec & pm$_\mathrm{RA}$ & pm$_\mathrm{Dec}$ & References \\
\hline
LMC & -18.1 &  $18.477\pm 0.026$  & 80.9 & -69.8 &  $1.85\pm 0.03$  &  $0.234\pm 0.03$ & $[1], [2], [4]$ \\ 
SMC & -16.5 &  $18.977\pm 0.028$  & 13.2 & -72.8 &  $0.797\pm 0.03$  &  $-1.22\pm 0.03$ & $[1], [2], [5]$ \\ 
Sag. & -13.5 &  $17.25\pm 0.008$  & 283.8 & -30.5 &  $-2.692\pm 0.001$  &  $-1.359\pm 0.001$ & $[1], [2], [6]$ \\ 
Fornax & -13.4 &  $20.77\pm 0.03$  & 40.0 & -34.4 &  $0.382\pm 0.001$  &  $-0.359\pm 0.002$ & $[1], [3], [7]$ \\ 
Leo I & -12.0 &  $22.07\pm 0.07$  & 152.1 & 12.3 &  $-0.05\pm 0.01$  &  $-0.11\pm 0.01$ & $[1], [3], [7]$ \\ 
Sculptor & -11.1 &  $19.62\pm 0.04$  & 15.0 & -33.7 &  $0.099\pm 0.002$  &  $-0.16\pm 0.002$ & $[1], [3], [8]$ \\ 
Leo II & -9.8 &  $21.68\pm 0.11$  & 168.4 & 22.2 &  $-0.14\pm 0.02$  &  $-0.12\pm 0.02$ & $[1], [3], [9]$ \\ 
Sextans & -9.3 &  $19.55\pm 0.01$  & 153.3 & -1.6 &  $-0.41\pm 0.01$  &  $0.04\pm 0.01$ & $[1], [3], [6]$  \\ 
Carina & -9.1 &  $20.12\pm 0.11$  & 100.4 & -51.0 &  $0.53\pm 0.01$  &  $0.12\pm 0.01$  & $[1], [3], [4]$ \\ 
Draco & -8.8 &  $19.35\pm 0.01$  & 260.1 & 57.9 &  $0.042\pm 0.005$  &  $-0.19\pm 0.01$ & $[1], [3], [6]$ \\ 
Ursa Min. & -8.8 &  $19.18\pm 0.02$  & 227.3 & 67.2 &  $-0.124\pm 0.004$  &  $0.078\pm 0.004$  & $[1], [3], [6]$
\end{tabular}}%
\end{table}

\begin{table} 
\centering 
\caption{Derived kinematical
quantities and their uncertainties for the Milky Way's
11 brightest satellites, ordered by V-band magnitude,
$M_V$. The quantity $r_\mathrm{GC}$ is the distance to the Galactic centre,
$r_\parallel$ is the distance from the centre of the plane within the
plane, $r_\perp$ is the distance from the centre of the plane
perpendicular to the plane, $v_\parallel$ is the speed relative to the
plane within the plane, $v_\perp$ is the speed relative to the plane
perpendicular to it. Positions and velocities are relative to the
centre of positions and in the rest frame of the plane. The orbital
period, $\tau$, and the apocentre radius, $r_a$, are in the default
potential with a halo mass of $10^{12}\Ms$. The errors reflect the
observational errors and are estimated using Monte Carlo sampling.
\label{tab:kinematics}}
\vspace{.5cm}
\resizebox{\columnwidth}{!}{%
\begin{tabular}{lllllllll}
 \hline\hline
 Name & $r_\mathrm{GC}$ & $r_\parallel$ & $r_\perp$ & $v_\parallel$ & $v_\perp$ &  $\tau$ & $r_a$ \\
  & [kpc] & [kpc] & [kpc] & [kms$^{-1}$] & [kms$^{-1}$] & [Gyr] & [kpc] \\
\hline   
LMC & $49.1 \pm 0.6$  & $50.9 \pm 1.1$  & $16.9 \pm 0.3$  & $237.1 \pm 4.8$  & $82.8 \pm 2.0$  & $1.64 \pm 0.11$  & $106.8 \pm 7.7$ \\ 
SMC & $59.7 \pm 0.7$  & $71.1 \pm 1.3$  & $26.8 \pm 0.4$  & $161.5 \pm 4.6$  & $100.7 \pm 1.8$  & $1.34 \pm 0.07$  & $71.4 \pm 4.6$ \\ 
Sagittarius & $20.4 \pm 0.1$  & $59.2 \pm 1.1$  & $21.7 \pm 0.2$  & $134.7 \pm 2.4$  & $265.4 \pm 2.0$  & $0.61 \pm 0.01$  & $47.2 \pm 0.9$ \\ 
Fornax & $144.5 \pm 2.0$  & $145.7 \pm 2.0$  & $37.3 \pm 0.5$  & $66.0 \pm 4.1$  & $57.8 \pm 2.5$  & $2.35 \pm 0.1$  & $149.4 \pm 2.4$ \\ 
Leo I & $262.7 \pm 8.9$  & $221.0 \pm 8.1$  & $15.5 \pm 0.4$  & $108.1 \pm 1.9$  & $77.8 \pm 1.5$  & $8.57 \pm 0.33$  & $501.0 \pm 18.8$ \\ 
Sculptor & $84.1 \pm 1.5$  & $106.5 \pm 1.8$  & $11.6 \pm 0.3$  & $227.1 \pm 1.9$  & $80.0 \pm 1.2$  & $1.51 \pm 0.03$  & $96.2 \pm 1.7$ \\ 
Leo II & $219.1 \pm 10.2$  & $187.4 \pm 9.3$  & $6.8 \pm 0.5$  & $56.8 \pm 5.6$  & $114.7 \pm 1.0$  & $3.74 \pm 0.2$  & $221.8 \pm 10.4$ \\ 
Sextans & $84.3 \pm 0.4$  & $41.5 \pm 1.3$  & $8.7 \pm 0.2$  & $63.9 \pm 1.0$  & $205.5 \pm 0.8$  & $1.93 \pm 0.02$  & $112.3 \pm 1.1$ \\ 
Carina & $108.0 \pm 5.7$  & $86.8 \pm 4.9$  & $14.0 \pm 0.5$  & $120.2 \pm 12.5$  & $25.5 \pm 4.6$  & $2.19 \pm 0.39$  & $110.0 \pm 11.9$ \\ 
Draco & $70.4 \pm 0.6$  & $82.3 \pm 0.8$  & $19.3 \pm 0.4$  & $228.1 \pm 1.4$  & $58.2 \pm 1.0$  & $1.11 \pm 0.01$  & $77.8 \pm 0.7$ \\ 
Ursa Minor & $74.1 \pm 0.3$  & $97.7 \pm 0.7$  & $11.3 \pm 0.5$  & $241.7 \pm 1.4$  & $3.9 \pm 1.0$  & $1.23 \pm 0.01$  & $86.1 \pm 0.5$ 
\end{tabular}}%
\end{table}

\clearpage

\clearpage

\end{document}